# Microstructure-Stabilized Blue Phase Liquid Crystals

*Jia-De Lin,[†,‡] Ying-Lung Daniel Ho,[‡,]\* Lifeng Chen,[‡] Martin Lopez-Garcia,[‡] Shun-An Jiang,[†] Mike P.C. Taverne,[‡] Chia-Rong Lee,[†,]\* and John G. Rarity[‡,]\**

† Department of Photonics, National Cheng Kung University, No.1, University Road, Tainan 701, Taiwan

‡ Department of Electrical and Electronic Engineering, University of Bristol, Merchant Venturers Building, Woodland Road, Bristol BS8 1UB, UK

ABSTRACT: We show that micron-scale two-dimensional (2D) honeycomb microwells can significantly improve the stability of blue phase liquid crystals (BPLCs). Polymeric microwells made by direct laser writing improve various features of the blue phase (BP) including a dramatic extension of stable temperature range and a large increase both in reflectivity and thermal 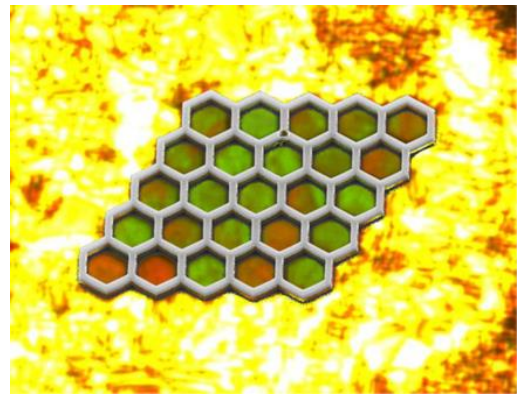 stability of the reflective peak wavelength. These results are mainly attributed to the omni-directional anchoring of the isotropically oriented BP molecules at the polymer walls of the hexagonal microwells and at the top and bottom substrates. This leads to an omni-directional stabilization of the entire BPLC system. This study not only provides a novel insight into the mechanism for the BP formation in the 2D microwell but also points to an improved route to stabilize BP using 2D microwell arrays.

■ INTRODUCTION

Photonic crystals (PCs) form naturally through self-ordering of spherical particles to create the iridescent colors in gem stones such as opal,[1] whereas the dynamic light manipulation properties of PCs are clearly seen in nature, for instance, in the vivid displays of cuttlefish.[2] Hence, the engineering of three-dimensional (3D) materials capable of actively manipulating light for novel functional devices has attracted much attention. Self-assembly methods have been pursued as an approach to engineer these PCs initially using soft matter materials, such as colloidal particles[3] using hard sphere packing[4, 5] and electrostatic repulsion[6] even achieving dynamic control through electric fields.[7] More recently polymers[8] and liquid crystals (LCs)[9] have been shown to self-organize into wavelength scale structures. In fact, a broad range of LCs with chirality possess a periodic structure and exhibit strong wavelength and polarisation-dependent scattering including chiral nematic phase (N*) or cholesteric LC (CLC) and blue phase (BP). The BP is a meso-phase between isotropic and chiral nematic phase, where first the LC molecules self-assemble into double-twisted cylinders (DTCs) with disclinations among adjacent DTCs, and then second these cylinders stack into woodpile-like structures with period of the order of the visible wavelength. By virtue of the adjustable periodicity and the specific chirality for a BP structure, a tunable photonic bandgap (PBG) for circular polarizations with handedness the same as that of the helix of the DTCs[10] is formed. This tunable PBG and its exceptional properties (e.g., submillisecond response time) open opportunities for the BP to have application in advanced photonic devices exploiting their 3D PC properties.[11] However, BP usually exists in a very narrow temperature range (a few degrees) because the formation of double-twisted helical structure is energetically unfavorable and critical.[12] To overcome this significant disadvantage, various techniques, such as polymer-stabilization,[13] nano-particle doping,[14-16] photosensitive chiral doping,[17] and molecular modification,[18] have been proposed for stabilizing the BP structure and widening its stable temperature range. The most common method to widen the temperature range of BP is polymer-stabilization. It has recently been reported that polymer-stabilized BP can exhibit a wide temperature range of more than 95 °C, leading to display applications.[19]

In this paper, we study the optical features and dynamic growth of the BP within a honeycomb microwell structure. Experimental results indicate that the BPs enclosed in the micro-honeycomb structure are significantly more stable than those out of the microwells. The BP temperature range on cooling in the microwell is dramatically increased (approximately fourfold) compared to BP outside the microwells. Additionally, the thermal stability of the peak wavelength and the reflectivity for the PBG in the microwell are both much improved. The results are attributable to the omnidirectional anchoring forces provided by the polymer walls of the microwell, and the sample substrates prevent the secondary logpile structure from dispersing once it has been formed, resulting in the significant improvement of optical features of the BP. This study not only provides a novel insight into the mechanism for the retention of BPs in two-dimensional (2D) microstructures but also demonstrates a new approach of developing a stabilized BP device for use in display devices or as an optical modulator.

## ■ RESULTS AND DISCUSSION

A scanning electron microscopy (SEM) top-view image of the honeycomb microstructure is shown in **Figure 1**. The minimum wall thickness and height are 5.4 and 6.4 $\mu$m, respectively, as shown in the inset of Figure 1. Each microwell cell in the honeycomb microstructure has a maximum inner width of around 20 $\mu$m. Figure 1b shows the microscopic 2D image of the fabricated honeycomb microwell array. For identification of hexagonal cells in the microstructure array, each cell is numbered as Area 1 to Area 25 while the area out of the microstructures is labeled Area 26, as shown in Figure 1b.

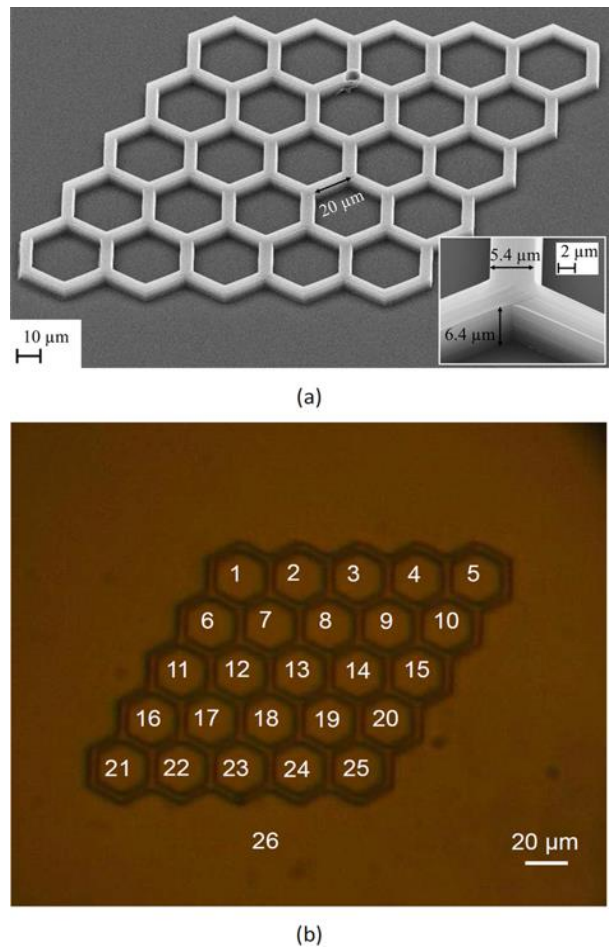

**Figure 1.** (a) SEM image of the micro-honeycomb structures fabricated using two-photon polymerization lithography. The minimum wall thickness is 5.4 $\mu$m (inset, scale bar, 2 $\mu$m), and the aspect ratio is 1.19 (height to wall thickness). The scale bar is 10 μm. (b) The microscopic image of the microstructure and the labeled areas, Area 1 to Area 25 in the microwells and Area 26 out of the micro-honeycomb structure. The scale bar is 20 $\mu$m.

After drop-casting the BP mixture on the microstructure substrate and enclosing it by directly placing a cover slip on the top of the substrate, the obtained BP-filled microstructure sample is observed by using a home-made transmission-type polarizing optical microscope (POM) equipped with crossed polarizers and a hot stage (LTS120, Linkam) and a temperature controller (T96, Linkam). The transmission-type, rather than reflection-type, POM is selected herein for the identification of the existence of BPs in the microwell cells. The colorful BP image can be easily observed through the transmission-type POM with crossed polarizers because the transmitted light with a certain color will dominantly leak through the optical activity of the BP and then the analyzer. In contrast the use of the reflection-type POM cannot assure the existence of BP if the reflection planes of the BP crystals are oblique relative to the observation direction. The sample was heated to 53 °C, which is above the clear point of the BP mixture, and then slowly cooled down at a rate of 0.1 °Cmin$^{-1}$. The POM images of the sample observed at different temperatures are shown in **Figure 2**. When the temperature is higher than 49 °C, the BP material is isotropic and thus only the microstructure can be seen under the POM with crossed polarizers (the image of an isotropic medium should be completely dark under the POM with crossed polarizers, but the exposure time of the camera was extended deliberately for clear observation). The crystallization of BP begins to appear in the region out of the microstructures at 48 °C, but at the same time only few areas in the microstructure exhibit BP. However crystallized regions grow across all areas both inside and outside the microstructures until all the observed regions in the microstructure are full of BP when the temperature reaches 46 °C. As the temperature falls to 44 °C, the LC outside of the microstructures (Area 26) begins to transition from BP to focal conic CLC, which appears as scattered white-yellow, whereas all areas inside the microstructure (Area 1 to Area 25) remain in the BP state until 38 °C. Below 38 °C, the LCs in more and more areas of the microstructure change from BP to focal conic texture however some survive down to 23 °C, where there are still four areas (Areas 1, 3, 6, and 19) where the LC remains in the BP state. When the temperature is cooled to 18.4 °C, the LCs in all areas of the microstructure transit to the focal conic state. The whole cooling process is recorded as Movie S1 in the Supporting Information.

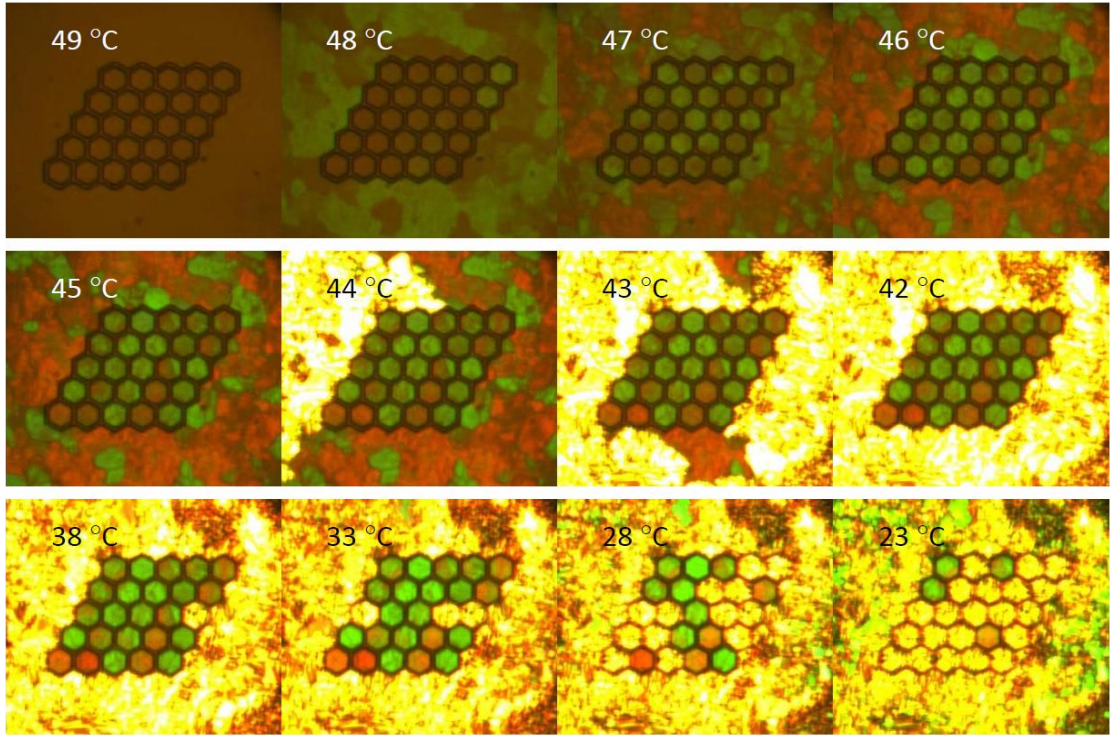

**Figure 2.** Transmission-type POM images of the BP sample with the honeycomb microwell structure at decreasing temperatures from 49 to 23 °C. The polarizer and analyzer in the POM are set as crossed.

The experimental results in the first-round cooling process in Figure 2 were used to calculate the variations in the BP temperature range and displayed in **Figure 3**a. We see random variations in the BP temperature range in different areas of the microstructure. To investigate if this is truly random or associated with specific microwells, we repeat the above-mentioned cooling process. The associated experimental results of the second-round are summarized in Figure 3b. By comparing results in Figure 3a,b, we see that there are random fluctuations in the temperature range of the BP in the microstructure but the enhancement of range is largely independent of the microwell number. The average temperature range of the BP in the microstructure is about 21 °C which is about four times of that outside of the microstructure in area 26 which is around 5 °C. We propose two main reasons for the different temperature range of BP in different areas. The primary reason is likely to be the slight variation in boundary conditions including the surface conditions of the substrates and the polymer wall of the microwells. A secondary reason might be slight variation of temperature across the sample. The extension of BP temperature range in each area is repeatable qualitatively but not quantitatively. As shown in Figure 3, the thermal properties of the BP are independent of the position of the microwells. Only one well, well 15, produces a consistently lower extension and even then it is twice the stability width of Area 26. We suspect there may be a small fabrication fault in well 15. A similar experiment with reduced honeycomb size with a unit cell of 10 μm width was performed, and the temperature range of BP in the microwells was also wider than that outside the microwells. We have not tried larger structures but expect to lose the effect if the unit cell is too large.

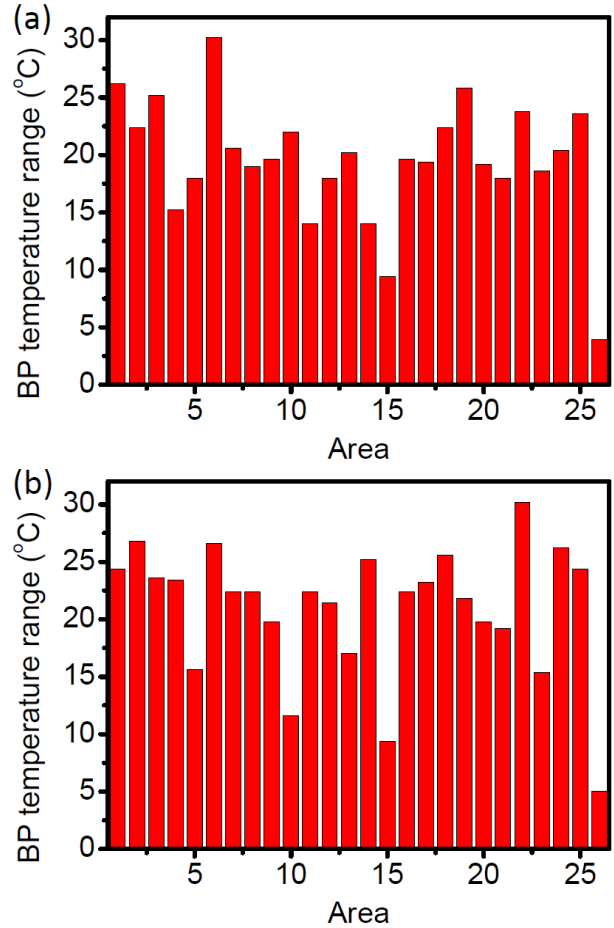

**Figure 3.** BP temperature range measured in all areas of the microstructure (Area 1 to Area 25). Area 26 is the area out of the microstructure. (a) First-round and (b) second-round cooling processes.

To compare the photonic properties of the BP inside and outside of the microstructures at different temperatures, we further measured the reflection spectra of four selected areas, three inside, in areas 2, 9, and 13, and the other outside in area 26. These are measured at a range of temperature during the slow cooling process (0.1 °C min$^{-1}$) and the associated data are summarized in **Figure 4**a−d, respectively. We have pre-confirmed that the relation between the optical properties of the BP with the temperature outside the microwells (Area 26) is identical to those of the BP sample without microstructure. The horizontal and perpendicular axes in the figures represent the temperature and wavelength, respectively, and the color bar from blue to red represents the reflectance of 0–50% in each measured reflection spectrum. The reflection colors of the BPs are determined by the reflecting lattice plane of the crystal, the lattice constant, and the plane normal relative to the observer. In this paper, the reflections at a central wavelength of ~550 nm from the crystal plane of (110) in the BPI crystal structure were measured. However, the other (200) and (211) planes leading to reflections at 389 and 318 nm have not yet been investigated. This is due to the fact that shorter wavelengths are out of the measurable range of the spectrometer. Other areas in the microwells show similar reflective properties to those shown in Areas 2, 9, and 13. The temperature-dependent Bragg reflection of the BP out of the microwells is shown in Figure 4d.

This result highlights the differences between the temperature-dependent photonic structures in the two areas, when we consider temperature range, reflectance, and the peak wavelength for the reflection band of the BP. First, as expected we see the temperature range of the BP ($\Delta T_{BP}$) in Areas 2, 9, and 13 of the microwells is considerably wider than that in Area 26. The isotropic-BP transition temperatures ($T_{I-BP}$) measured in both areas are similar (between 48 and 49°C) but the BP–CLC transition temperatures ($T_{BP-CLC}$) measured in Areas 2, 9, 13 and 26 are, respectively, approximately 28, 24, 26 and 42°C. That is, the temperature range to sustain in the BP state in Area 9 is about 4 times wider than that in Area 26 consistent with the results shown in Figure 2 and 3. The result shows that the LCs tend to remain in the BP state rather than reverting to the focal conic state under the influence of the microstructure during cooling. Second, both the average reflectance in Area 9 over the entire BP temperature range and the maximum reflectance (approximately 35 and 49%, respectively) are much higher than those in Area 26 (approximately 22 and 35%, respectively). The higher reflectivity of BP in the microwell reveals the fact that the BP is partially oriented by the walls of the microwell and thus is closer to a single crystal.[20] Third, the peak wavelength of the reflection in Area 9 is much more thermally stable than that in Area 26. The reflection band and peak wavelength in both areas have a similar trend on cooling showing an initial blue-shift followed by red-shift just after the LC enters the BP state below $T_{I-BP}$. The small blue-shift, concurrently with a low reflectance, can be explained by a slightly unstable obliqueness of the reflection planes of the BP crystals at the initial forming stage of the platelets around the $T_{I-BP}$ and the later red-shift concurrently with an increased reflectance arises from the intrinsic property of negative temperature-dependence of pitch for a general BP system (i.e., $dP/dT < 0$). After the red-shift, the peak wavelength of reflection in Area 9 remains constant around 558 nm for approximately 20 °C, wheres Area 26 effectively shows no stable temperature region. The peak stability over 20 °C of the BP state in Area 9 suggests high stability of the formed BP crystals leading to high reflectivity over the entire range. All the abovementioned results provide clear evidence that the structure of the honeycomb microwells combined with the substrate boundary provides an omnidirectional stabilization of the BP crystals significantly improving the photonic features. The lowest energy state or stable state of LC can change when LC is constrained by external microstructures[21] as evidenced by previous work, which showed similar effects in nematic LC using polymeric microscaffolds to extend the metastable state.[22] This supports our observation of the large extension of the temperature range of the BP inside microstructures presented in this work. These manifest improvements of the BP stability in microstructures opens up potential for applications outside the laboratory, for example in displays and photonic switches.

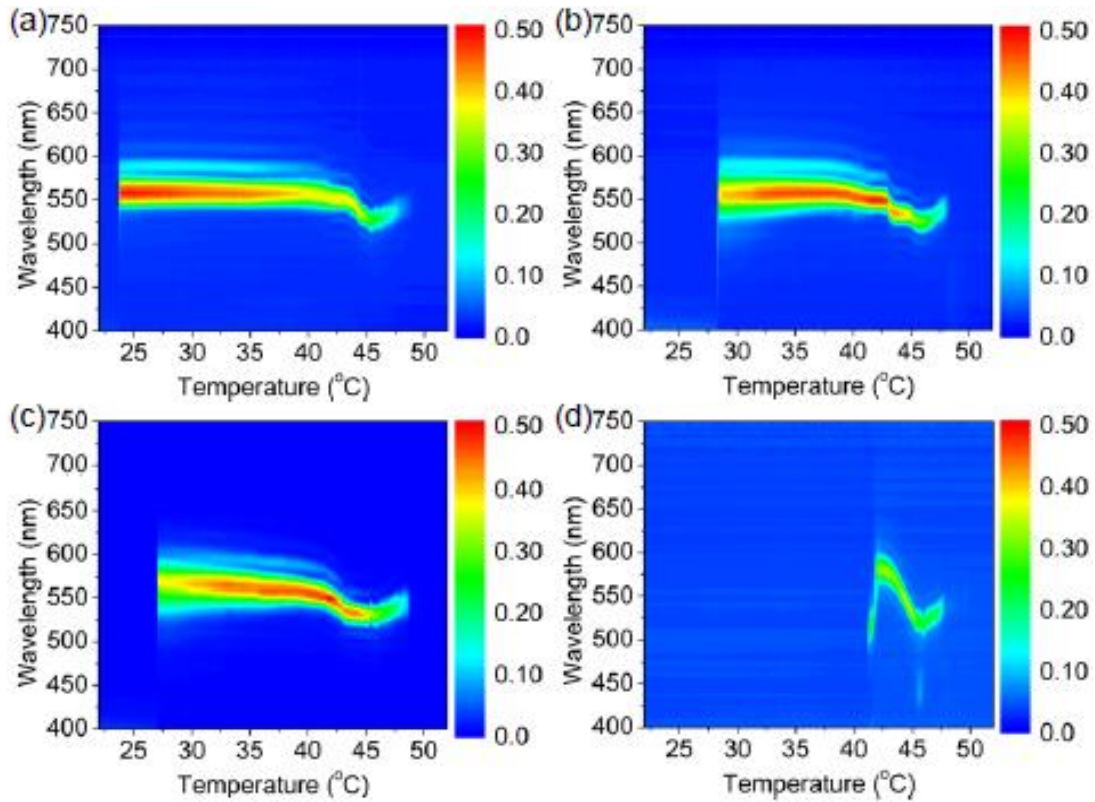

**Figure 4.** Temperature-dependent reflection spectra of the BP in (a) Area 9, (b) Area 2, (c) Area 13, and (d) Area 26 in and out of the honeycomb microwells, respectively.

To verify the above-mentioned omnidirectional stabilization effect of the sample with the honeycomb microwells, we perform similar cooling experiments based on two traditional BP samples with the same BP mixture as that used in the abovementioned experiment but without the microstructure. The two samples are divided into four regions with different boundary conditions: regions 1 and 2 with nonrubbed and rubbed polyvinyl alcohol (PVA) film, respectively, and 3 and 4 with nonrubbed and rubbed polyimide (PI) film, respectively. Each rubbed sample was mechanically rubbed three times. The transition points of $T_{\text{I-BP}}$ and $T_{\text{BP-CLC}}$ and $\Delta T_{\text{BP}}$ at different regions of the two samples are recorded and summarized in Table 1. The values of $T_{\text{I-BP}}$ and $T_{\text{BP-CLC}}$ measured in the four regions of the BP samples are limited between 50 and 52 °C and between 43 and 45 °C, respectively, such that the values of $\Delta T_{\text{BP}}$ lie between 5 and 7 °C. This result shows that the discrepancy of the BP temperature range for the four regions of different alignment conditions is very small. According to the previous literature,[23–26] the general order for the azimuthal anchoring strength in regions 1–4 ($W_i$, $i$ = 1–4) would be $W_1 \approx 10^{-7}$ to $10^{-6}$ J/m$^2$ < $W_2 \approx 10^{-5}$ to $10^{-4}$ J/m$^2 \cong W_3 \approx 10^{-5}$ to $10^{-4}$ J/m$^2$ < $W_4 \approx 10^{-4}$ to $10^{-3}$ J/m$^2$. Such a large difference of azimuthal anchoring strength, for example, between $W_1$ (isotropically weak anchor) and $W_4$ (anisotropically strong anchor) can only lead to a very small difference of $\Delta T_{\text{BP}}$ less than 1.0 °C. This result shows that both the strength and directionality of the anchoring force arising from the sample substrate alone has only a small impact on the transition points and only effectively stabilizes the blue phase liquid crystal (BPLC) molecules with orientations parallel to the substrate plane. As the BP ordering involves large scale 3D structure it seems logical that stabilization in other directions will be necessary to maintain a BP as the lowest free energy state.

**Table 1. Transition Points of $T_{\text{I-BP}}$ and $T_{\text{BP-CLC}}$ and $\Delta T_{\text{BP}}$ at Different Regions of the Two Samples with Different Alignment Films (PVA and PI) and Different Surface Treatments (Nonrubbed and Rubbed) are Recorded and Summarized.**

|  | with nonrubbed PVA | with rubbed PVA | with nonrubbed PI | with rubbed PI |
|---|---|---|---|---|
| $T_{\text{I-BP}}$ (°C) | 51.7 | 50.9 | 50.1 | 50.4 |
| $T_{\text{BP-CLC}}$ (°C) | 46.7 | 45.3 | 43.0 | 44.5 |
| $\Delta T_{\text{BP}}$ (°C) | 5.0 | 5.6 | 7.1 | 5.9 |

In general, it is necessary to apply an electric field or provide an anisotropically strong anchoring force on the LC molecules from the sample substrates during the formation of the BP crystals to obtain more uniformly oriented BP crystals.[27–29] Herein, although no electric field or strong anchoring force from the aligned substrates is exerted on the LC molecules, the results suggest that the hexagonal micro-cell can lead to highly ordered crystal growth and high uniformity of lattice plane orientation. In other words, the BPs in each microwell of the microstructure have a higher probability to grow into a single-crystal-like domain, unlike the polycrystalline BP outside the microstructure. However, whether or not highly ordered crystal growth can contribute to a wider BP temperature range, was not fully investigated. Further studies will be performed to determine the relation between the condition for crystal growth and the temperature range of BP.

Previous literature has demonstrated that two kinds of nucleation process probably occur in the crystal growth of BP: homogeneous and heterogeneous nucleations,[30, 31] corresponding to nucleation in the bulk and at the surface, respectively. We expect homogeneous nucleation to lead to randomly ordered and oriented polycrystalline BP whereas heterogeneous growth from walls of the microstructure should lead to the growth of ordered single crystals. To more clearly identify the mechanism for the process of the crystal growth of the BP in the honeycomb microwell cell, we focused the illumination light into a single unit of the microwell to dynamically observe the reflective image of BP in a single hexagonal microcell as it cooled from the isotropic state to room temperature. The process is recorded in Movie S2 (refer to the Supporting Information) and the images at some important moments of the movie are selected and summarized in **Figure 5**. It should be noticed that the concentration of the chiral dopant used in the experiment is reduced slightly (the concentration of the chiral dopant becomes 32.5 wt %) to make more apparent color change during the crystal growth process. When the sample is cooled just below the clear point, the twisting force from the chiral dopant begins to orient the LC molecules in a helical sense locally, and thus randomly orientated DTCs form uniformly in the entire microcell. These can be observed from the blue-green scattering of these DTCs, as shown in Figure 5a. The bright blue-green DTCs tend to redistribute and aggregate to the center and the six corners of the hexagonal micro-cell [Figure 5b]. With further decrease of temperature, the DTCs at some corners of the hexagonal microcell begin to stack with each other in an ordered way through interaction with the polymer walls of the corners. In the heterogeneous nucleation stage, red cubic BP crystals with the crystalline planes normal to the substrate near these corners form, as shown in Figure 5c. These ordered BP crystals then grow continuously from the corners to the center of the microcell [Figure 5d] and eventually cover almost the entire microcell (the area near the right corner of the micro-well is over-exposed in the camera). The red BP crystals remain in a wide temperature range under the stabilization of the polymer walls of the microcell until the temperature approaches $T_{BP-CLC}$. Once arriving at $T_{BP-CLC}$, the reflection color in the center of the microcell starts to become emerald green [Figure 5e] and then diffuses to the boundary of the microcell concurrently with the blue-shift to the final blue color as the temperature is lower than $T_{BP-CLC}$ [Figure 5f, g]. From Figure 5e–g, we find that the BP near the boundary of the honeycomb can be retained in the BP state slightly longer before turning into a focal conic state during the natural cooling process. This result is reasonable given the stabilization from the polymer walls is weakest at the center. The transient blue-shift of the reflection color is probably attributed to the reorganization of the chiral dopants from 3D orientation in BP with the double-twist structure to 2D orientation in the CLC phase with the single-twist structure. Figure 5h is the final stage of the cooling process, at which the LC material in the microstructure has become the stable CLC focal conic phase completely.

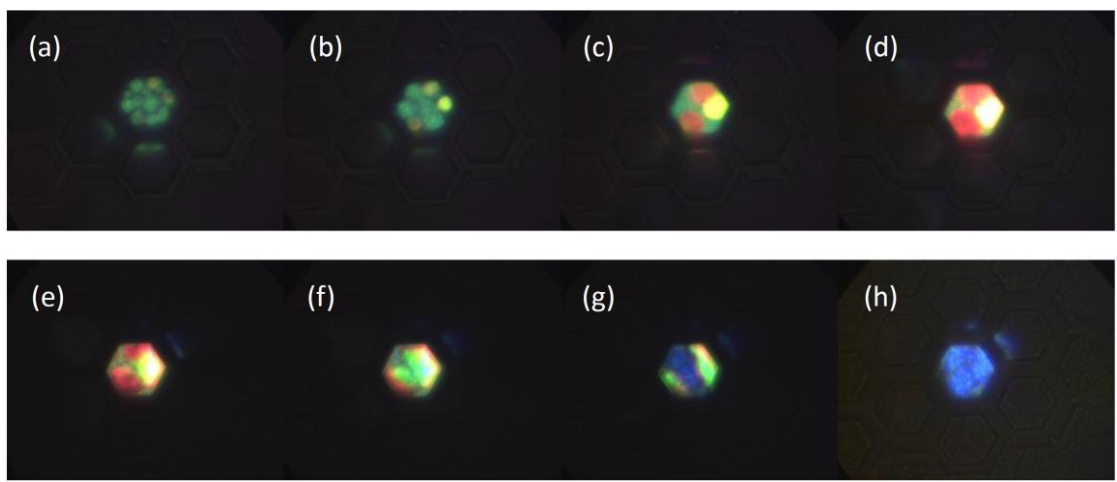

**Figure 5.** Developing process of the BP in the microstructure in free cooling experiment (~0.1 °C/s). (a) Initially uniform formation and (b) redistribution of the DTCs in the hexagonal microcell [(a, b) were recorded at 49–47 °C]. (c) Heterogeneous nucleation of ordered BP via the stack of the DTCs near the corners and (d) growth of the BP crystals [(c, d) were recorded at 47–30 °C]. (e)−(g) Transition of the BP with the double-twisted structure to CLC with the single-twisted structure from the center to the boundary of the cell, concurrently with the blue-shift of reflection color [(e−g) were recorded at around 30 °C]. (h) Final CLC focal conic state.

The data in Figure 5 suggest that the polymer walls of the microcell can stabilize the BP crystal structure for an extended temperature range. We have done preliminary tests to see if the stability is the same on warming and, so far, we see little or no extension of the BP stability range. This supports the previous work[32] which suggests that the PC DTC rod structure is metastable and locked in place by the microstructure once it has formed. However, the microstructure does not change the temperature, (~48 °C) for the onset of BP formation.

■ CONCLUSION

In summary, we fabricated a micro-honeycomb structure by DLW and found it can significantly improve several optical features of the BP including a dramatic extension of the stable temperature range, much higher reflectivity, and much better thermal stability of the reflection peak wavelength. The main factor contributing to this result is the weak boundary force provided by the polymeric microwell structure combined with the substrate of the sample which helps to anchor and template the growth of uniformly oriented DTC logpile crystals. Once these templated crystals have formed it becomes difficult for the DTC to unwind into the CLC phase under the stabilization of the polymer walls. This statement is supported by warming experiments and other studies[25] where we find little or no extension of the BP stability region. This microstructure stabilization of the BP also triggers new inspiration about the academic research into BP-based devices and their further industrial applications.

■ EXPERIMENTAL SECTION

**Sample Preparation.** The 2D honeycomb microwell structures in this paper were fabricated using a commercial DLW system (Photonic Professional, Nanoscribe GmbH) based on the two-photon polymerization (TPP) technique. A 780 nm femtosecond laser beam (with pulse width ≈ 120 fs and repetition rate ≈ 80 MHz) was focused into a negative photoresist (IP–L 780, Nanoscribe GmbH) for triggering TPP process by a high numerical aperture (NA = 1.4) oil immersion objective lens (100×, Zeiss). There were no specific procedure requirements regarding cleaning the microwells. The developed polymer was rinsed in isopropyl alcohol and then dried in clean nitrogen. Following this, the BP mixture was drop-cast onto the microstructure filling the array. Then a coverslip without rubbing treatment was directly placed on the top of the microstructure to enclose the BP mixture. The height of the microstructure, 6.4 $\mu$m, is the thickness of the LC layer. The BP mixture used in this study was composed of nematic LC, HTW-114200-100 (Fusol), and chiral dopant, S811 (Merck), in the ratio of 64.6:35.4 by weight. The material was first heated to the isotropic state and injected into the 2D honeycomb microwell structures on the substrate by drop-cast. During the injecting process, the temperature of the substrate was controlled at 53 °C by a hot stage for ensuring the isotropic state of the BP mixture. After each hexagonal cell in the 2D honeycomb microwell structures was full of the mixture, the excess of the mixture was removed and then a cover slip is put on the top of the microwell for enclosing it. Following the sample was cooled down with a slow cooling rate of 0.1 °Cmin$^{-1}$ for the growth of the BP crystals in the microstructure cells.

**Experimental Setups.** Home-made transmission-type POM was employed for observing the phases of the LC material in the sample of microstructure array. For measuring reflectance, a setup was built up separately as described below. Light from a halogen bulb was coupled to a multimode fiber (with a core diameter of 300 μm) and collimated by an objective lens (4×). The collimated light was focused by a high numerical aperture (magnification: 63×, NA: 0.75) objective lens on the sample to reduce the spot size around 20 μm, which was smaller than the unit cell of the microstructure array. Reflected light was collected by the same objective lens and picked off in a 50% beamsplitter followed by an achromatic lens which refocus the reflection light to a detection plane, either a CMOS camera (Thorlab DCC1645C) for imaging, or a detecting fiber with a core diameter of 200 μm and connecting to a spectrometer (USB2000, Ocean Optics) for spectrometry.

■ ASSOCIATED CONTENT

**Supporting Information**

The Supporting Information is available free of charge on the ACS Publications website at DOI: 10.1021/acsome-ga.8b01749.

> Whole cooling process of the BP in the 2D honeycomb microwell structures is recorded as Movie S1 (ZIP)
> 
> Crystalline growth process of BP in one microwell is recorded as Movie S2 (ZIP)

■ AUTHOR INFORMATION

**Corresponding Authors**
*E-mail: Daniel.Ho@bristol.ac.uk (Y.-L.D.H.).
*E-mail: crlee@mail.ncku.edu.tw (C.-R.L.).
*E-mail: John.Rarity@bristol.ac.uk (J.G.R.).

**ORCID**
Jia-De Lin: 0000-0002-5361-5530
Ying-Lung Daniel Ho: 0000-0001-8643-4990
Chia-Rong Lee: 0000-0002-7917-7583

**Author Contributions**

J.-D.L. and Y.-L.D.H. conceived the idea. L.C., M.P.C.T, and Y.-L.D.H fabricated the microstructure. J.D.L. and S.-A.J. prepared the sample. J.D.L., L.C., M.L.-G., and S.-A.J. carried out the experiment. All the authors discussed the results and contributed to the final manuscript. Y.-L.D.H., C.-R.L. and J.G.R. supervised the project.

**Notes**

The authors declare no competing financial interest.

■ ACKNOWLEDGEMENT

J.G.R. and Y.-L.D.H. acknowledge financial support from the ERC advanced grant 247462 QUOWSS, EPSRC grants EP/M024458/1 & EP/M009033/1, and the Royal Society − International Exchanges Award IE161622. C.-R.L. and J.-D.L. acknowledge the Ministry of Science and Technology of Taiwan (contract numbers: MOST 103-2917-I-006-003 and MOST 106-2911-I-006-509) for financially supporting this research.